\documentclass[11pt]{article}
\usepackage{moriond}

\def\be{\begin{equation}}
\def\ee{\end{equation}}
\def\bea{\begin{eqnarray}}
\def\eea{\end{eqnarray}}

\def\fermi{\emph{Fermi}}
\newcommand{\Photo}{}

\begin{document}
\vspace*{4cm}
\title{THE DISCOVERY OF A POPULATION OF $\gamma$-RAY NOVAE}

\author{ A.B. HILL$^{1,2}$, C.C. CHEUNG$^{3}$ \& P. JEAN$^{4}$ on behalf of the \fermi-LAT collaboration}

\address{
\ \\$^{1}$W. W. Hansen Experimental Physics Laboratory, KIPAC, Department of Physics and SLAC National Accelerator Laboratory, Stanford University, Stanford, CA 94305, USA\\
$^{2}$Faculty of Physical \& Applied Sciences, University of Southampton, Southampton, SO17 1BJ, UK\\
$^{3}$Space Science Division, Naval Research Laboratory, Washington, DC 20375, USA\\
$^{4}$Institut de Recherche en Astrophysique et Plan\'{e}tologie, 9 av. colonel Roche, BP 44 346, 31028 Toulouse Cedex 4, France
}

\maketitle\abstracts{
Novae have long been expected to be sources of emission at several MeV from the decay of radioactive elements in the novae ejecta, however, they were not anticipated to be sources of continuum emission in the GeV energy domain. In March 2010 the Large Area Telescope (LAT) on-board the \fermi\ {\em Gamma-ray Space Telescope} discovered for the first time $>$100 MeV gamma-ray emission from a nova within our galaxy, V407 Cyg.  The high-energy spectrum and light curve was explained as a consequence of shock acceleration in the nova shell as it interacts with the local ambient medium.  While this was an exciting and important discovery it was suspected that the necessary conditions for high-energy emission from novae would be rare.  In June 2012 the LAT detected two new transient sources that have been associated with classical novae observed in the optical, Nova Sco 2012 and Nova Mon 2012.  We report on the observational properties of the population of gamma-ray novae, their similarities and differences and the emission processes that generate the high energy radiation in these systems.}

\section{Introduction}
Novae have been observed by astronomers for thousands of years although it is only in more modern times that the processes behind these cosmic eruptions has become clear.  Classical novae (CNe) are a subset of cataclysmic variables, binary systems that host a white dwarf (WD) that accretes material from the secondary star.  In the vast majority of these systems the orbital period ranges between $\sim$1.4--8 hours and the secondary star is a low mass, main sequence star; there are a few longer period systems in which for the secondary star to fill its Roche Lobe it has to have evolved off the main sequence\cite{Bode}.

The WD accretes material off the secondary star building up H-rich material on its surface.  Once sufficient material has accumulated for the critical pressure/temperature to be achieved at the base of the accreted envelope then a thermonuclear runaway explosion is triggered producing a CN event.  The outburst increases the luminosity to L$\sim$10$^{4}$L$_{\odot}$ and large amounts of mass are ejected at high velocities\cite{Bode}, typically M$_{ejecta}\sim10^{-5}-10^{-4}$~M$_\odot$ and v$_{ejecta}\sim10^{2}-10^{3}$~km s$^{-1}$. The inter-outburst period of CN explosions is believed to be $\sim10^3-10^4$ years. 

Conversely, recurrent novae (RNe) have inter burst timescales of the order of decades.  Only ten such systems have been identified within our Galaxy and are broken into three subclasses:
RS Oph/T Crb systems with long orbital periods ($\sim$100s days) and red giant donors with heavy winds; U Sco systems are characterised by containing a more evolved main sequence or sub-giant secondary with orbital periods of order a day and extremely high ejection velocities; T Pyx are short orbital period systems that show slower optical decays than the other RNe.

MeV $\gamma$-ray line emission has long been predicted from novae as a consequence of the decay of radioactive elements produced in the nova explosion\cite{Hernanz}, however, prior to the discovery of GeV emission from V407 Cyg by the \fermi\ Large Area Telescope (LAT) in 2010\cite{ref3} there had been no predictions of continuum emission at such high energies.

\begin{figure}[t]
\centering
\includegraphics[width=130mm]{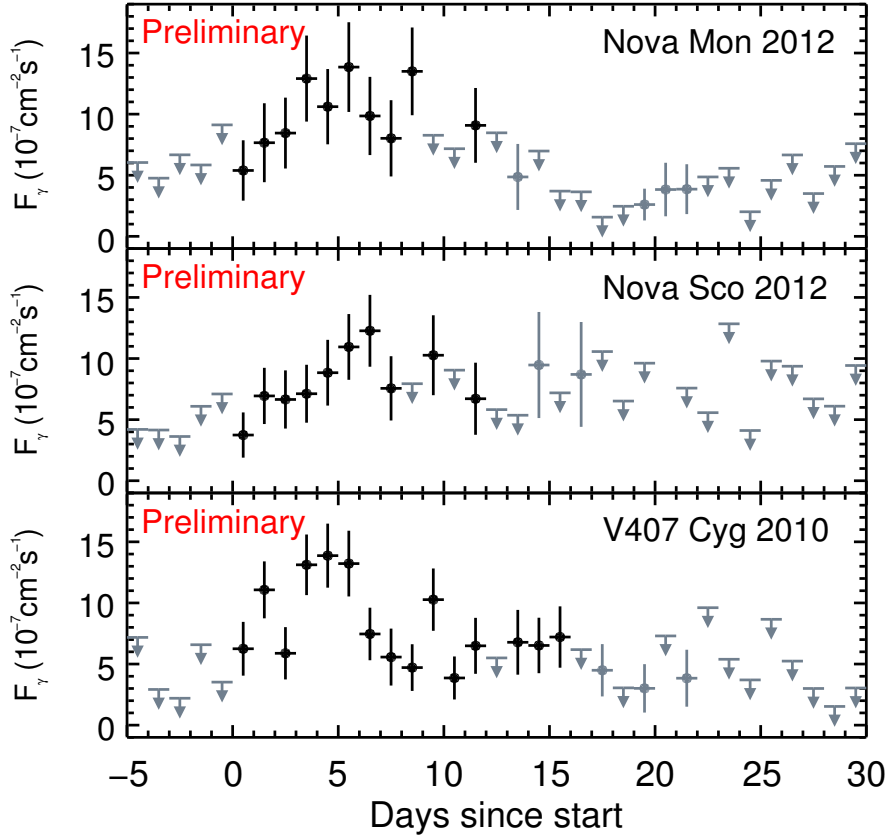}
\caption{\fermi-LAT 1-day binned ($>100$ MeV) light curves of the three detected 
$\gamma$-ray novae. The start dates indicate the i$\gamma$-ray onset and are from top to bottom, 2012 June 9, 
2012 June 15, and 2010 March 10. Detections with $\geq 3\sigma$ significances 
are the points with flux error bars shown in black, while data with 2--3$\sigma$ 
significances are shown in gray; upper limits are shown for points with 
$<2\sigma$ significances.}
\label{fig1}
\end{figure}

\section{V407 Cyg: The first $\gamma$-ray nova}
On 11 March 2010 Japanese amateur astronomers reported the discovery of a new 8$^{th}$ magnitude nova in the Cygnus constellation \cite{ref1}.  The nova was identified as originating from the known symbiotic binary, V407 Cyg.  This systems comprises of a hot WD accreting from a Mira-type variable red giant and consequently the WD is embedded in a particularly dusty environment generated by the heavy wind of the donor star.

The discovery of a classical nova event in this system was completely unexpected compounded by the unexpected discovery of $\gamma$-ray emission above 100 MeV from the nova by the LAT\cite{ref2,ref3}. The $\gamma$-rays were detectable for approximately two weeks after the optical nova onset with an average spectral energy distribution in the form of a power law with an exponential cutoff and flux above 100 MeV of $4.4 \times 10^{-7}$ ph cm$^{-2}$ s$^{-1}$.  The onset of $\gamma$-ray emission was consistent with the optical onset.  The $\gamma$-ray  light curve of the 2010 V407 Cyg eruption is shown in the lower panel of Figure~\ref{fig1}.  

The $\gamma$-ray emission was explained by the nova ejecta shell expanding outwards and colliding with the stellar wind of the red giant companion and forming a shock front where particles could be accelerated to high energies.  The line connecting the WD with the donor star contains the largest local density enhancement attributable to the red giant wind in which the nova shell can sweep up material in a fashion similar to what is ascribed in supernova explosion models.  Taking the estimated parameters of the system (wind density, ejecta velocity, etc...) and the measured spectral energy distribution, it was shown that this model could feasibly produce $\gamma$-ray emission through the decay of neutral pions produced in proton collisions or inverse Compton (IC) scattering off accelerated electrons\cite{ref3}. More detailed modelling of the environment has suggested that leptonic processes may be dominant\cite{Pierrick}.  The direct link between the dense local wind environment and the $\gamma$-ray production mechanism and the rarity of symbiotic and RS Oph type RNe systems led to the suggestion that `$\gamma$-ray' novae would be exceptionally rare events\cite{Nelson}.

\begin{table}[t]
\caption{A comparison of some of the properties of V407 Cyg, Nova Sco 2012 and Nova Mon 2012.}
\begin{center}
\begin{tabular}{l|ccc}
\hline
System & V407 Cyg\cite{ref3}   & Nova Sco 2012\cite{ref7,Hill} & Nova Mon 2012\cite{Shore,KDwarf}\\
\hline
Optical companion  & Mira-type red giant & ? & K Dwarf?\\
Distance & 2.7 kpc & $\sim$4--5 kpc & 3.6 kpc\\
Nova spectral class & He/N& Fe-II & Neon nova: ONe WD \\
Speed class & Very fast & Fast/moderately fast & Fast?\\
& t$_{2} \sim$5.9 days & t$_{2} \sim$25 days & ? \\
$\gamma$-ray spectrum & Cutoff power law & Power law & Power law\\
$\gamma$-ray duration & $\sim$16 days & $\sim$12 days & $\sim$12 days\\
Optical/$\gamma$-ray delay & $<$3 days & $\sim$14 days & ?\\
Peak $\gamma$-ray flux & $\sim$1.4 &  $\sim$1.3 & $\sim$1.4\\
$\times 10^{-6}$ ph cm$^{-2}$ s$^{-1}$ & \\
\hline
\end{tabular}
\end{center}
\label{pop}
\end{table}

\section{A new Galactic transient: Fermi J1750$-$3243}
From 16--30 June 2012 \fermi\ identified a new $\gamma$-ray source, Fermi J1750$-$3243, that was not consistent with any of the known 2FGL catalog sources \cite{ref5}.  This new source was localised to RA = 267.727$^\circ$, Dec = $-$32.720$^\circ$ with a 95\% error radius of 0.122$^\circ$ \cite{ref4}.  The location of the new LAT transient was consistent with the report of a newly discovered optical nova, MOA 2012 BLG$-$320 (Nova Sco 2012) which had entered into optical outburst between June 1.77--2.15 2012 when it brightened dramatically in the I band from 17$^{th}$ magnitude to 11$^{th}$  magnitude \cite{ref6}. Subsequent IR spectral observations on June 17.879 indicated that it was an Fe-II nova event with an ejecta velocity of $\sim$2,200 km s$^{-1}$ \cite{ref7}.  It appeared that \fermi\ had discovered another `$\gamma$-ray nova'.  However this system appeared to more like traditional CNe systems with no indication of a dense stellar environment and it's behaviour at other wavelengths was quite different to that observed in V407 Cyg\cite{Hill}.

\section{Nova Mon 2012: A nova first discovered in $\gamma$-rays}
On 22 June 2012 another new unidentified $\gamma$-ray source was detected by the LAT, Fermi J0639$+$058\cite{NovaMon}; proximity to the Sun prohibited follow-up observations at other wavelengths.  In August 2012 a late-stage optical nova, Nova Mon 2012, was discovered within the LAT error circle\cite{NovaMon2}. The first optical spectra were taken approximately 55 days after the $\gamma$-ray peak and identified the WD as being a member of the ONe class\cite{Shore} i.e. the more massive class of WDs.  This system has also been associated with the more common CNe systems and has been identified as having a tight binary orbital period of $\sim$7.1 hours\cite{7.1hr}.

\section{Summary}
The \fermi-LAT light curves of all three novae are shown in Figure~\ref{fig1} and can be seen to be remarkably similar.  However, as demonstrated by the summary of some their more general multi-wavelength characteristics in Table~\ref{pop} there are also a number of distinct differences.  Most notable is that the two 2012 novae appear to be CNe in the traditional sense and so the $\gamma$-ray production mechanism invoked to explain the emission in V407 Cyg cannot be applied suggesting that there is another mechanism in action in these systems.  This raises the potential for many other CNe and RNe to be $\gamma$-ray bright, although it is equally evident that numerous other optically detected CNe have not been reported by that LAT.  Only with further study and analysis will we be able to identify the $\gamma$-ray production channels and determine what fraction of the Galactic novae population are capable of producing emission at these high energies.

\section*{Acknowledgments}
A.~B. Hill acknowledges that this research was supported by a Marie Curie International Outgoing Fellowship within the 7th European Community Framework Programme (FP7/2007--2013) under grant agreement no. 275861. The $Fermi$ LAT Collaboration acknowledges support from a number of agencies and institutes for both development and the operation of the LAT as well as scientific data analysis. These include NASA and DOE in the United States, CEA/Irfu and IN2P3/CNRS in France, ASI and INFN in Italy, MEXT, KEK, and JAXA in Japan, and the K.~A.~Wallenberg Foundation, the Swedish Research Council and the National Space Board in Sweden. Additional support from INAF in Italy and CNES in France for science analysis during the operations phase is also gratefully acknowledged.

\end{document}